\documentclass[jmp,amsmath,amssymb,preprint]{revtex4-1}
\usepackage{dcolumn}
\usepackage{bm}
\usepackage{color}
\usepackage{rotating}
\usepackage{graphicx}
\usepackage{latexsym}
\usepackage{amssymb}
\usepackage{amsfonts}
\usepackage{soul}
\usepackage{color}
\usepackage{amsmath}
\usepackage[utf8]{inputenc}
\usepackage[T5]{fontenc}
\usepackage[main=english]{babel}
\usepackage{csquotes}
\usepackage{comment}
\usepackage{tikz}
\usepackage{pifont}
\usepackage{amssymb,amsmath,bm,epsfig,graphicx,subfigure,hyperref}
\usepackage[most]{tcolorbox}

\newtcbox{\caphl}{
  colback=yellow,
  colframe=yellow,
  boxrule=0pt,
  arc=0pt,
  left=2pt,
  right=2pt,
  top=1pt,
  bottom=1pt,
  boxsep=0pt
}
\usepackage[utf8]{inputenc}
\usepackage{chngcntr}
\usepackage{xcolor}
\usepackage{braket}
\usepackage{physics}
\usepackage{sidecap}
\usepackage{lipsum}
\usepackage{amsfonts}
\usepackage{bbold}
\usepackage{tensor}
\usepackage{makecell}
\usepackage{multirow}
\usepackage{rotating}
\usepackage{float}
\usepackage{amstext}

\newif\ifshowfullcontent
\showfullcontenttrue

\hypersetup{
    colorlinks=false,
    citecolor=black,
    linkcolor=black,
    urlcolor=black,
    pdfborder={0 0 0}
}
\usepackage{lineno}
\begin{document}

\title{Family of High-Chern-Number Orbital Magnets in Twisted Rhombohedral Graphene}

\author{Xirui Wang$^{1}$}
\email{xrwang@mit.edu}
\thanks{These authors contributed equally to this work}
\author{L. Antonio Ben\'itez$^{1}$}
\email{lbenitez@mit.edu}
\thanks{These authors contributed equally to this work}
\author{Skandaprasad Rao$^{1}$}
\author{Võ Tiến Phong$^{2,3}$}
\author{Wai In Chu$^{1}$}
\author{Kenji Watanabe$^{4}$}
\author{Takashi Taniguchi$^{5}$}
\author{Cyprian Lewandowski$^{2,3}$}
\author{Pablo Jarillo-Herrero$^{1}$}
\email{pjarillo@mit.edu}
\affiliation{$^1$Department of Physics, Massachusetts Institute of Technology, Cambridge, Massachusetts 02139, USA. }
\affiliation{$^2$Department of Physics, Florida State University, Tallahassee, Florida, 32306, USA.}
\affiliation{$^3$National High Magnetic Field Laboratory, Tallahassee, Florida, 32310, USA.}
\affiliation{$^4$Research Center for Electronic and Optical Materials, National Institute for Materials Science, 1-1 Namiki, Tsukuba 305-0044, Japan.}
\affiliation{$^5$Research Center for Materials Nanoarchitectonics, National Institute for Materials Science, 1-1 Namiki, Tsukuba 305-0044, Japan.}

\begin{abstract}{
\textbf{
Realizing Chern insulators with Chern numbers greater than one remains a major goal in quantum materials research. Such platforms promise multichannel dissipationless chiral transport and access to correlated phases beyond the conventional $C = 1$ paradigm. Here, we discover a family of high-Chern-number orbital magnets in twisted monolayer–multilayer rhombohedral graphene, denoted $(1+n)$ with $n =$ 3, 4, and 5. Magnetotransport measurements show pronounced anomalous Hall effects at one and three electrons per moiré unit cell when they are polarized away from the moiré interface. Across the $(1+n)$ systems, we observe a clear topological hierarchy $C =  n$, revealed by the Středa trajectories and the quantized Hall resistance. Our experimental observations are supported by self-consistent mean-field calculations. Moreover, we realize both electrical and magnetic switching of the high-Chern-number states by flipping the valley polarization. Together, these results establish a tunable hierarchy of orbital Chern magnets in twisted rhombohedral graphene, offering systematic control of Chern number and topology through layer engineering in pristine graphene moiré systems.
}
}

\end{abstract}


\maketitle

Quantized Hall conductance at zero magnetic field arises when the Berry curvature integrated over the Brillouin zone yields a nonzero, integer-valued Chern number $C$. This bulk topological invariant guarantees the existence of chiral edge states, which in turn produce conductance quantized in $C$ multiples of the conductance quantum. The resulting phase is known as a Chern insulator. Realizing Chern insulators beyond the conventional $C= 1$ regime opens access to topological phases with multiple co-propagating chiral edge modes, providing a direct route to multi-channel, low-dissipation transport. Moreover, high-Chern-number systems are predicted to host a hierarchy of exotic fractional and non-Abelian excitations~\cite{Wang2012Fractional,Barkeshli2012Topological,Liu2012Fractional,Sterdyniak2013Sterdyniak,Liu2013Fractional, moller2015fractional,andrews2018stability,andrews2021stability}, which emerge from the interplay between strong correlations and topology, and are central to the pursuit to fault-tolerant topological quantum computation \cite{RevModPhys.80.1083}. Motivated by this, magnetic topological insulators have shown that $C$ can be tuned by magnetic doping or layer stacking, thereby establishing a practical design rule for realizing higher-order quantum anomalous Hall states~\cite{zhao2020tuning}.

In parallel, van der Waals heterostructures offer an exciting and versatile platform, with rhombohedral graphene being a particularly promising material candidate, whose valley Chern number increases systematically with the number of layers~\cite{zhou2021half,han2024correlated,liu2024spontaneous,han2024large,sha2024observation}. Moreover, theoretical studies predict that the interlayer hybridization between two twisted rhombohedral multilayers enables the design of Chern bands with tunable topology via displacement field and twist angle, as well as allowing the Chern number to be selected by appropriate layer combinations~\cite{ledwith2022family,wang2022hierarchy,phong2025coulomb}. This approach provides a powerful route to design and control topological band structures, yet experimental realizations of such high-Chern-number moiré systems have only recently begun to emerge~\cite{dong2025observation,wang2025moir}. Importantly, these systems can go beyond the graphene–hBN moiré paradigm, where the broken inversion symmetry in hBN has a strong influence on the electronic properties of rhombohedral graphene but the exact mechanism remains incompletely understood~\cite{uzan2025hbn}. 

In this work, we engineer a family of twisted monolayer–multilayer rhombohedral graphene (multi-RhG) structures, denoted $(1+n)$, to systematically study how the Chern number evolves as we increase the number of layers. By applying a small twist angle between monolayer and multi-RhG flakes from the same parent crystal, we create a well-defined moiré potential at the interface, enabling controlled exploration and engineering of topological band structures.

\section{Band structure evolution with displacement field}

\begin{figure}[ht!]
    \centering
\includegraphics[width=1\linewidth]{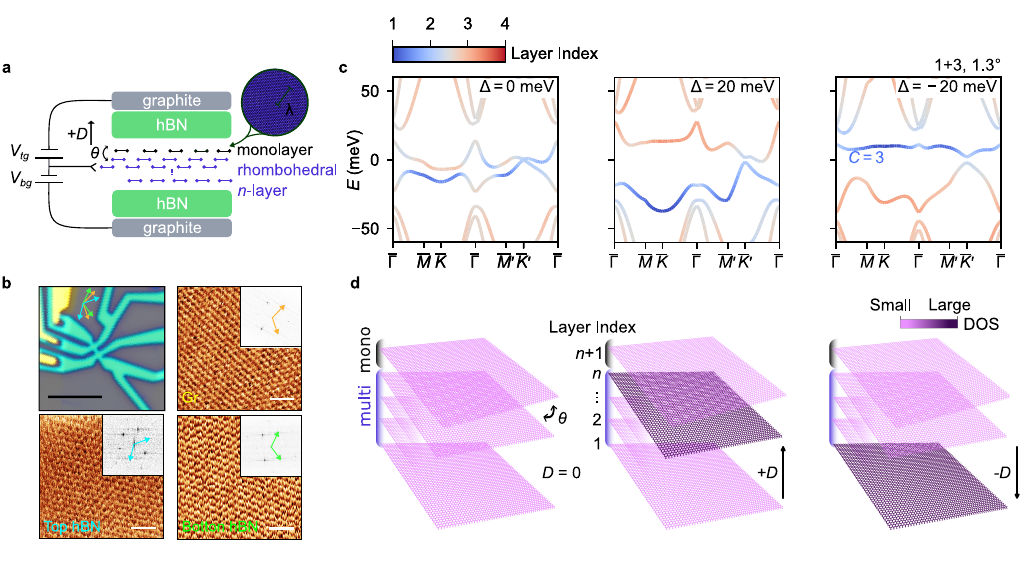}
    \caption{\textbf{Device structure, moiré characterization, and band structure calculations in twisted $(1+n)$ systems.} \textbf{a,} Schematic of a twisted $(1+n)$ heterostructure, consisting of a monolayer graphene placed on an $n$-layer rhombohedral graphene with a twist angle $\theta$, forming a moiré superlattice of wavelength $\lambda$. Top and bottom graphite gates enable independent control of carrier density $n$ and displacement field $D$.
\textbf{b,} Optical micrograph (top left) of a typical device used in this study, and lateral force microscopy (LFM) images of graphene (Gr) and top and bottom hexagonal boron nitride (hBN) lattices. The Fourier transform insets reveal the threefold or sixfold symmetry of the atomic lattices, with armchair directions indicated. The scale bars correspond to 5 µm in the optical image and 1 nm in the LFM images.
\textbf{c,} Single-particle band structures of twisted $(1+3)$ graphene at $\theta = 1.3^\circ$ at interlayer potential differences $\Delta$ of 0, 20, and $-20$ meV. At $\Delta=-20$ meV, the first conduction band has $C=3$. The color represents layer polarization, with 4-red (1-blue) indicating charge localization on the monolayer (multilayer away from moiré interface) side. 
\textbf{d,} Schematic illustration of the evolution of layer polarization with $D$, showing electronic states localized near the moiré interface for $D>0$ and shifted toward the bottom multilayer surface for $D<0$. DOS: Density of states.}
    \label{fig:1}
\end{figure}

The device schematic and an optical image of a representative device are shown in Fig.~\ref{fig:1}a,b. The device consists of a multi-RhG with a monolayer graphene (Gr) placed on top, forming a twist angle $\theta$ and thereby generating a moiré pattern characterized by moiré wavelength $\lambda$. We encapsulate the $(1+n)$ stack with hBN dielectrics, and intentionally misalign them with the graphene lattice to prevent the formation of secondary moiré structures. To this end, we perform lateral force microscopy (LFM) with atomic resolution on both the graphene and hBN flakes. For example, the graphene is misaligned by approximately 30° and 20° from top and bottom hBN, respectively, as shown in Fig.~\ref{fig:1}b for a twisted (1+3) device. Outside the hBN layers, we use top and bottom graphite gates to independently control the displacement field $D$ and carrier density $n$. We confirm the rhombohedral stacking order of the multilayer graphene by Raman mapping (see Methods and Extended Data~\ref{fig:ED2} and~\ref{fig:ED3}), before patterning the stack into a Hall-bar geometry for measuring the longitudinal ($R_\mathrm{xx}$) and transverse ($R_\mathrm{yx}$) resistances.

According to the single-particle band structure calculations in Fig.~\ref{fig:1}c, in the absence of a perpendicular displacement field $D$, the low-energy states of the $(1+n)$ system are intrinsically layer-unpolarized. The monolayer graphene is labeled as layer $n+1$, and the layer index of multi-RhG stack is denoted by $1,2,...,n$ ( $n$ = 3, 4, or 5). For $\Delta = -20~\text{meV}$ ($D < 0$), the first conduction band flattens and becomes nearly fully polarized toward the bottom layer of the multi-RhG, while carrying a nontrivial Chern number ($C \neq 0$). The calculation in Fig.~\ref{fig:1}c corresponds to a $(1+3)$ system with a twist angle of $1.3^\circ$, though a similar behavior occurs for $(1+3)$, $(1+4)$ and $(1+5)$ structures, predicting a Chern number of 3, 4 and 5 at moderate negative displacement fields, respectively (see Extended Data~\ref{fig:ED9} and Supplementary Figs. S1-S3). 
The strong electronic interactions due to the flatness of the conduction band are expected to split the band and generate a cascade of correlated phases (see Supplementary Information). At $\Delta = 20~\text{meV}$ ($D > 0$), the first conduction band wave function becomes polarized toward the moiré interface. The pronounced band dispersion around the $\overline{\Gamma}$ point indicates that the kinetic energy dominates, thereby suppressing electron–electron correlation effects. By varying $D$, one can thus continuously drive the system between distinct electronic configurations, ranging from flat, layer-polarized bands with non-trivial topology and strong interaction effects, to more dispersive, weakly correlated regimes. Figure~\ref{fig:1}d schematically illustrates this evolution, highlighting the position of the wave functions either close to the moiré interface for $D > 0$ or farther from it for $D < 0$. In our calculations, we treat $\Delta$ only as a \textit{proxy} for $D$ that is input into the non-interacting continuum Hamiltonian, while an exact mapping between the two would require a self-consistent electrostatic calculation. The relation between $\Delta$ and $D$ is explained in the Supplementary Information.

\section{High-Chern-number quantum anomalous Hall effect in twisted \boldmath$(1+n)$\unboldmath\ graphene}

\begin{figure}[ht!]
    \centering
    \includegraphics[width=1\linewidth]{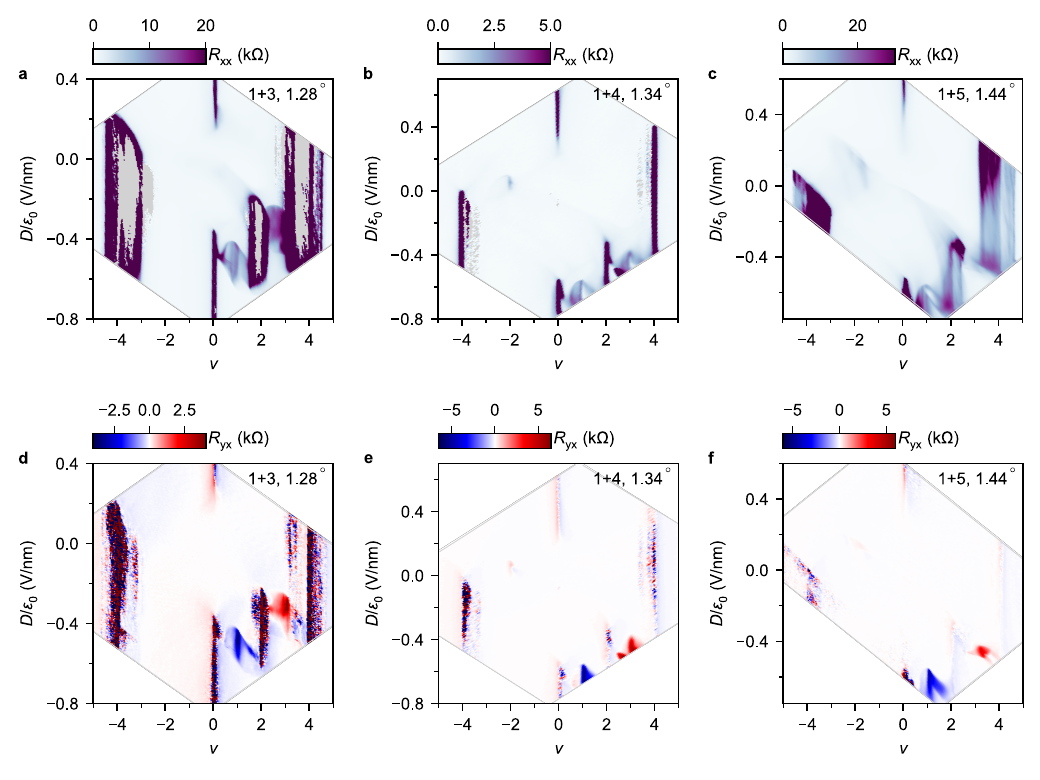}
   \caption{\textbf{Longitudinal and transverse resistance maps of $(1+n)$ systems.} \textbf{a-c,} Longitudinal resistance $R_\mathrm{xx}$ as a function of filling factor $\nu$ and displacement field $D/\varepsilon_0$ for $(1+3)$, $(1+4)$, and $(1+5)$ devices with twist angles of $1.28^\circ$, $1.34^\circ$, and $1.44^\circ$, respectively. All the data of $(1+3)$ and $(1+4)$ devices in the main figures are from the same devices as in \textbf{a,b}. \textbf{d-f,} Corresponding dual-gate maps of transverse resistance $R_\mathrm{yx}$ for the same devices. Data were taken at out-of-plane magnetic field \textit{B} = 0.1 T, 0.15 T, and 0.06 T for $(1+3)$, $(1+4)$, and $(1+5)$ devices, respectively, with $R_\mathrm{xx}$ data symmetrized and $R_\mathrm{yx}$ antisymmetrized with respect to \textit{B}. All data were taken at \textit{T} = 0.3 K.}
    \label{fig:2}
\end{figure}

To reveal the electronic behavior of the $(1+n)$ systems, Fig.~\ref{fig:2} presents $R_\mathrm{xx}$ and $R_\mathrm{yx}$ maps as functions of the filling factor $\nu$ and the displacement field $D$, for $n = 3, 4,$ and $5$ (information and images of all the devices in this study are shown in Extended Data~\ref{tab:ED1} and Extended Data~\ref{fig:ED1}). Throughout this work, $R_\mathrm{xx}$ and $R_\mathrm{yx}$ denote the symmetrized and antisymmetrized components of the raw resistance data with respect to magnetic field, unless otherwise specified. $R_\mathrm{xx}$ exhibits insulating behavior at full fillings of conduction and valence bands, reflecting the band insulator formed from the moiré band folding. Near charge neutrality, the system develops a gap as $D$ increases, while on the hole side, $R_\mathrm{xx}$ remains metallic for all $D$, consistent with our band structure calculations (Fig.~\ref{fig:1}c). On the electron side, and particularly for negative $D$ which corresponds to states away from the moiré interface, the $R_\mathrm{xx}$ maps display pronounced resistive peaks close to integer fillings $\nu = 1$, 2, and 3, indicating tunable symmetry breaking and strong correlation effects. The corresponding $R_\mathrm{yx}$ maps (Fig.~\ref{fig:2}d-f) show clear $R_\mathrm{yx}$ hot spots near $\nu = 1$ and $\nu = 3$ for $D < 0$. Notably, the hot spot near $\nu = 1$ is highly reproducible across the different $(1+n)$ systems. Around $\nu = 3$ for the $(1+3)$ device, $R_\mathrm{yx}$ exhibits sign reversals as either $D$ or $\nu$ is tuned, indicating a doping- and displacement-field-controlled reversal of the sign of the Berry curvature, which will be discussed later on in detail.

To gain further insight into the $R_\mathrm{yx}$ hot spots, Fig. \ref{fig:3}a-c present detailed maps of $R_\mathrm{yx}$ for electron doping and $D < 0$. To verify their magnetic character, Fig. \ref{fig:3}d-f show $R_\mathrm{xx}$ (blue) and $R_\mathrm{yx}$ (black) as functions of the out-of-plane magnetic field $B$ at selected values of $\nu$ and $D$. At $\nu \approx 1$, the $(1+3)$, $(1+4)$, and $(1+5)$ devices exhibit similar responses with prominent anomalous Hall signals as well as hysteresis as \textit{B} is scanned forward and backward. In the $(1+4)$ and $(1+5)$ devices, $R_\mathrm{yx}$ develop  quantized plateaus corresponding to $C = 4$ and $C = 5$, respectively, indicative of the high-Chern-number quantum anomalous Hall state. A minimal $R_\mathrm{xx}$ of 0.2 k$\Omega$ is reached for the (1+4) device at quantization. Detailed measurements of the (1+5) device reveal that the quantization extends from $D = -0.63$ V/nm to $-0.57$ V/nm at $\nu = 1.0$, and a minimal $R_\mathrm{xx}$ of 3.5 k$\Omega$ is reached (Fig. S4 in the Supplementary Information). The achieved quantized $R_\mathrm{yx}$ is repeatable across multiple thermal cycles as shown in the Supplementary Information. 

Besides quantization at $C=4$ observed at $\nu=1$, $R_\mathrm{yx}$ of around 9.7 k$\Omega$ is observed at $\nu=2.73$ in the $(1+4)$ device, suggesting a Chern state with smaller Chern number. In the $(1+5)$ device, we observe $R_\mathrm{yx}$ hot spots between $\nu=1$ and $\nu=2$ that disperse with both $\nu$ and $D$, resembling features previously reported in rhombohedral graphene aligned with hBN~\cite{zheng2024switchable}. 

\section{Electrical and magnetic switching of high-Chern-number orbital magnets}

To reveal the underlying band topology, we performed measurements at higher magnetic fields. In general, the Hall resistance plateau of an incompressible Chern insulator follows the dispersion $C = \frac{h}{e}\frac{\partial n}{\partial B}$ according to the Středa relation~\cite{streda1982quantised}, which links the change in carrier density with magnetic field to the Chern number $C$. In the twisted $(1+3)$ device, the Hall plateau at $\nu = 1$ shifts with magnetic field with a slope corresponding to $C = 3$ as shown in Fig.~\ref{fig:4}a (dotted line as a guide to the eye). A line cut along this $C = 3$ trajectory is presented in Extended Data~\ref{fig:ED4}, showing that $R_\mathrm{yx}$ begins to saturate near 4~T at a value close to $|R_\mathrm{yx}| = h/3e^2$ (slightly higher). The fitting of the dispersion of $R_\mathrm{xx}$ minimum as a function of \textit{B} also yields a Chern slope of 3 (Extended Data~\ref{fig:ED4}), consistent with $C=3$ predicted by our calculations (Fig.~\ref{fig:1}c).  This observation indicates that electrons are polarized into a single valley as a result of strong electron-electron interactions. Another $(1+3)$ device exhibiting quantized $R_\mathrm{yx}$ around $\nu=3$ near zero \textit{B} is shown in Extended Data~\ref{fig:ED5}. In addition to the integer fillings, we also observe an $R_\mathrm{yx}$ hot spot emerging near $\nu = 1.5$, following a dispersion consistent with a Chern number $C = 2$ (Fig.~\ref{fig:4}a), pointing toward the possibility of exotic correlated phases such as a topological charge density wave, which combines broken translational symmetry with nontrivial topology~\cite{polshyn2022topological,xie2021fractional}. While a quantitative Středa analysis of $R_\mathrm{xx}$ near $\nu = 1.5$ is challenging in the current sample due to the proximity to the $\nu = 2$ correlation peak, this observation highlights the system as an intriguing platform for future exploration of interaction-driven phases. 

\begin{figure}[ht!]
    \centering
    \includegraphics[width=1\linewidth]{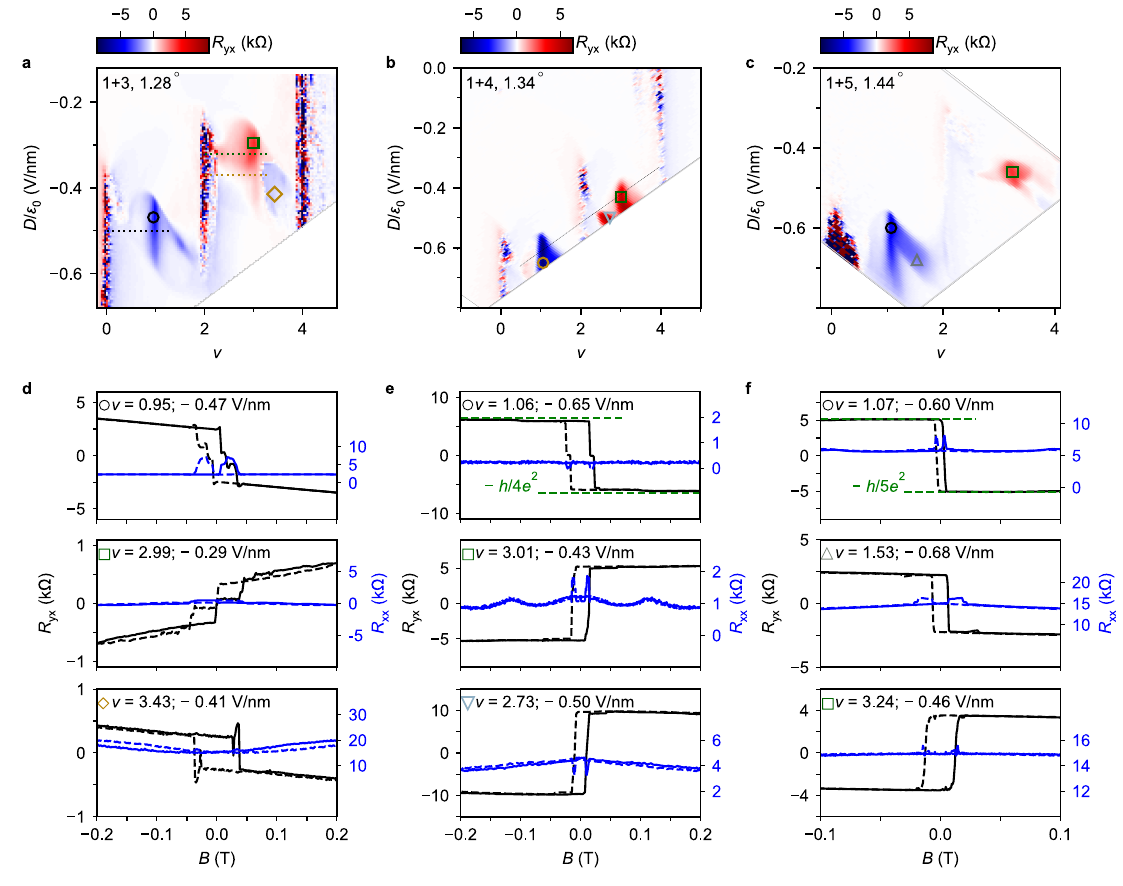}
     \caption{\textbf{Anomalous Hall and high-Chern-number quantum anomalous Hall effects.} \textbf{a-c,} Zoomed-in dual-gate maps of $R_\mathrm{yx}$ as a function of $\nu$ and $D/\varepsilon_0$ for $(1+3)$, $(1+4)$, and $(1+5)$ devices. \textbf{d-f,} $R_\mathrm{yx}$ (antisymmetrized, black) and $R_\mathrm{xx}$ (symmetrized, blue) measured as a function of $B$ for selected values of $\nu$ and $D/\varepsilon_0$, which are labeled with corresponding markers in \textbf{a-c}. Solid (dashed) lines denote forward (backward) \textit{B} sweeps. All data were taken at \textit{T} = 0.3 K.} 
    \label{fig:3}
\end{figure}

\begin{figure}
    \centering
    \includegraphics[width=1\linewidth]{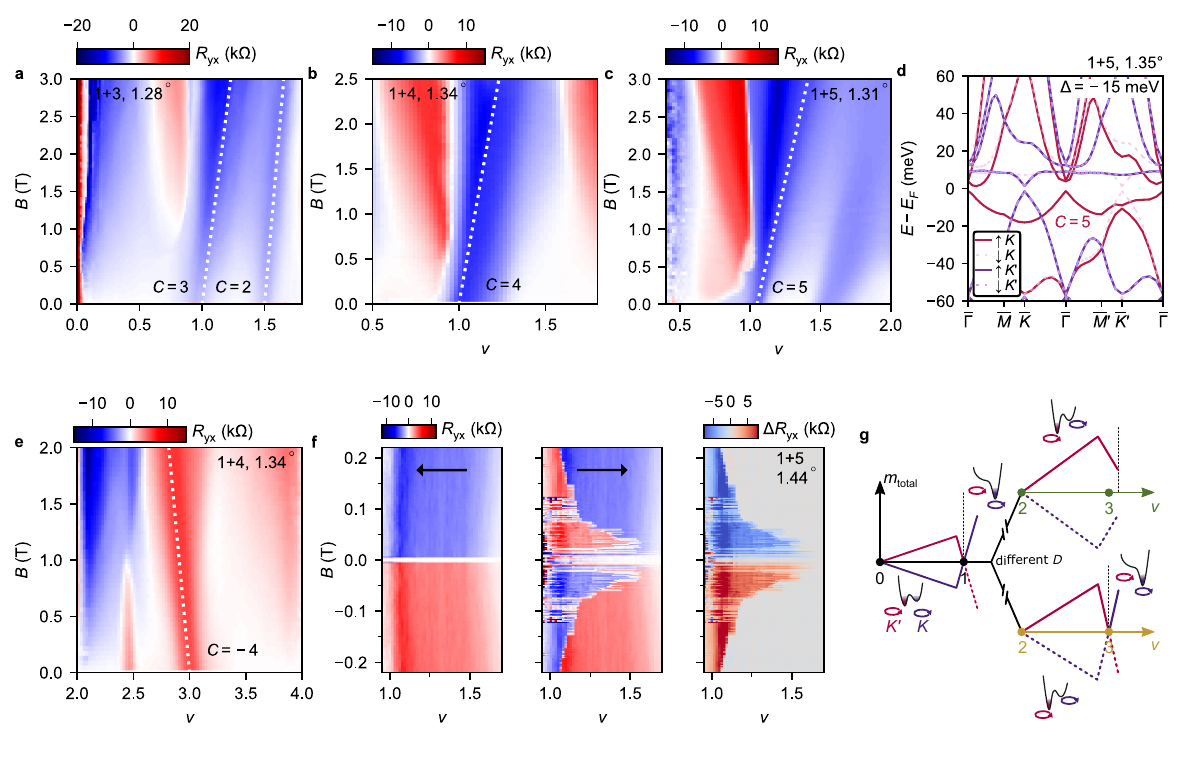}
\setlength{\abovecaptionskip}{-40pt}
\caption{\textbf{Electrical and magnetic switching of high-Chern-number orbital magnets. } $R_\mathrm{yx}$ as a function of $\nu$ and \textit{B} for $(1+3)$ device at $D/\varepsilon_0$ = $-0.53$ V/nm (\textbf{a}), for $(1+4)$ device (\textbf{b,e}), and for $(1+5)$ device at $D/\varepsilon_0 = -0.66$ V/nm (\textbf{c}). Due to gate limit, \textbf{b} and \textbf{e} are taken at constant top gate voltage. Corresponding linecut is noted with dashed line in Fig. 3b. The dotted lines which correspond to the noted Chern numbers are drawn as a guide to the eye. \textit{T} = 0.3 K. \textbf{d,} Hartree-Fock calculated band structure of $1.35^\circ$-$(1+5)$ graphene.  \textbf{f,} Doping-induced magnetization switching in $(1+5)$ device at $D/\varepsilon_0$ = $-0.64$ V/nm. $\nu$ scanned backward (left), forward (middle), and the difference between forward and backward scans (right). \textit{T} = 0.05 K. \textbf{g,} Schematic of doping-induced valley polarization switching. Red (purple) solid lines denote \textit{K'} (\textit{K}) valley occupied, and dashed lines denote unoccupied valleys. \textit{x}-axis: $\nu$. \textit{y}-axis: total magnetization $m_\mathrm{total}$ of either valley. The three segments denote linecuts along $\nu$ at different displacement fields (linecuts drawn in Fig. 3a with corresponding colors as an example).}
    \label{fig:4}
\end{figure}

In the $(1+4)$ device, besides quantization at zero field, linear fitting of the dispersion of  $R_\mathrm{yx}$ maximum and $R_\mathrm{xx}$ minimum yields a Středa slope of $C = 4$ from $\nu=1$ (Fig.~\ref{fig:4}b and Extended Data~\ref{fig:ED10}), agreeing with the theoretical prediction (Extended Data~\ref{fig:ED9}). At $\nu = 3$, as shown in Fig.~\ref{fig:4}e, the $R_\mathrm{yx}$ hot spot disperses along $C = -4$, indicating that the electrons now occupy the opposite valley compared to $\nu = 1$. In a $(1+5)$ device, we observe the $R_\mathrm{yx}$ dispersion emanating from $\nu = 1$ with a slope corresponding to $C = 5$, as shown in Fig.~\ref{fig:4}c, in line with Chern slope extracted from the dispersion of $R_\mathrm{xx}$ minimum (Extended Data \ref{fig:ED7}). At magnetic fields above 0.5~T in Fig.~\ref{fig:4}c, an additional dispersing feature appears, which likely originates from conventional Landau-level formation. Another $R_\mathrm{yx}$ hot spot dispersing from $\nu \approx 1.5$ corresponds to the hot-spot region between $\nu=1$ and $\nu=2$ observed in Fig.~\ref{fig:2}f and Fig.~\ref{fig:3}c. The Arrhenius fitting yields a thermal activation gap of 19.4 K for the (1+4) device and 18.3 K for the (1+5) device (Extended Data \ref{fig:ED8}). We note that incorporating electron–electron interactions within a self-consistent Hartree–Fock framework yields a spin–valley–polarized state preserving the same topological character as in the single-particle description (Fig.~\ref{fig:4}d and Extended Data~\ref{fig:ED10}). The calculated band structure further reveals that the electronic states are localized in layers distant from the moiré interface (Extended Data~\ref{fig:ED9}), similar to the behavior predicted in the twisted $(1+3)$ device. 

Having demonstrated that the twisted (1+\textit{n}) system exhibits a $C = n$ hierarchy, we now illustrate the electrical switching of the high-Chern magnetization. As shown in Fig.~\ref{fig:4}f, when we sweep the carrier density in the forward direction from $\nu=1$, $R_\mathrm{yx}$ switches sign, and a similar behavior is observed at around $\nu = 3$ in twisted $(1+3)$ and ($1+4$) samples (Fig.~\ref{fig:3}a and Extended Data~\ref{fig:ED4} and~\ref{fig:ED6}). Given the small spin magnetization in moiré systems, we attribute the doping-induced $R_\mathrm{yx}$ sign switching to competition between bulk and edge orbital currents, a mechanism previously invoked to explain electrical magnetization switching in moiré graphene systems~\cite{polshyn2020electrical,zhu2020voltage,tschirhart2021imaging,grover2022chern}. The bulk contribution comes from self-rotation of the wave packet which evolves within the band and is not directly probed by transport, while the edge contribution is from chiral edge currents when the Fermi level lies within a Chern gap and can be measured through transport. In a single valley, bulk and edge magnetization are opposite in sign  ~\cite{polshyn2020electrical,zhu2020voltage,tschirhart2021imaging,grover2022chern}.

Since $K$ and $K'$ valleys have opposite Berry curvature, a flip of the $R_\mathrm{yx}$ sign corresponds to the change of polarization between the two valleys. We examine three representative linecuts with different sequences of $R_\mathrm{yx}$ sign reversals using the schematic in Fig.~\ref{fig:4}g. The corresponding examples are denoted as linecuts in the dual-gate map of the ($1+3$) device (Fig.~\ref{fig:3}a). As the doping evolves from $\nu=0$ or 2 towards $\nu=1$ or 3, $R_\mathrm{yx}$ either stays negative (black line), positive (green line), or change from positive to negative (yellow line). The total magnetization $m_\mathrm{total}$ from either $K'$ (red) or $K$ (purple) valley is plotted as a function of doping. 

The system tends to minimize the Zeeman energy by aligning the total magnetization with the external magnetic field. When the Fermi level goes across the Chern gap, if the total magnetization of a single valley changes sign, the system tends to switch valley to maintain low Zeeman energy, leading to a reversal of $R_{\mathrm{yx}}$ from positive to negative (yellow line). In the quantized $(1+5)$ device, this enables electrical switching of the high-Chern quantization, as shown in Extended Data~\ref{fig:ED7}. On the other hand, if the sign of the total magnetization of a single valley remains the same (green line), it results in a continuous evolution of positive $R_\mathrm{yx}$ without sign reversal. The black line corresponds to the case where the large negative $R_\mathrm{yx}$ at $\nu=1$ indicates polarization into the \textit{K} valley, while the small negative $R_\mathrm{yx}$ at $\nu<1$ reflects nearly equal valley polarization, likely due to the large energy barrier for single-valley polarization at lower doping (Fig.~\ref{fig:4}g).

The above framework also explains the field-dependent sign reversal, where a high magnetic field tilts the energy landscape between $K$ and $K'$ valleys, so that a sign reversal as function of doping can happen which otherwise does not occur at small \textit{B}, as shown in Fig.~\ref{fig:4}a-c. Doping-induced valley switching is accompanied by pronounced hysteresis as carrier density is swept forward and backward (Fig.~\ref{fig:4}f, Extended Data~\ref{fig:ED4} and~\ref{fig:ED7}). This indicates that the valley switching occurs via a first-order phase transition probably due to domain-wall pinning between regions harboring opposite valley polarization. The divergence of the coercive magnetic field around $\nu = 1$ or 3 further suggests a minimal energy splitting between the $K$ and $K'$ valleys~\cite{tschirhart2021imaging}. We want to note that, while the above discussion provides a possible picture of the electrical magnetization switching mechanism, local imaging methods will enable a more definitive identification of its microscopic origin.

\section{Conclusion}

Our experiments establish the twisted $(1+n)$ family of graphene moiré systems as a promising platform for engineering high-Chern-number bands through controlled combinations of rhombohedral layer number and twist angle, with high reproducibility across multiple devices (Extended Data~\ref{fig:ED4},~\ref{fig:ED5}, and~\ref{fig:ED7}). A key and unique aspect of our work is that we propose and observe, for the first time, a systematic and predictable approach to engineering high-Chern-number orbital magnets in graphene moiré systems. While recent studies have explored individual moiré structures, often accompanied by theoretical descriptions developed independently for each specific system~\cite{dong2025observation,wang2025moir,liu2025diverse}, our work reveals a simple and unified design principle: for the $(1+n)$ family, the isolated Chern band carries Chern number $C=n$, giving rise to anomalous Hall states at fillings of one and three electrons per moiré unit cell. Despite the substantial differences among tri-, tetra-, and penta-layer rhombohedral graphene, their behavior when interfaced with monolayer graphene is remarkably similar: all exhibit robust, and in some cases quantized, anomalous Hall effects when electrons localize in the layer farthest from the moiré interface. Beyond the systematic Chern hierarchy, the observation of electrically switchable high-Chern quantization further highlights the robustness of this design principle and opens new opportunities for topological electronics.

It is instructive to compare our results with multi-RhG aligned to hBN. In that case, alignment produces an isolated flat Chern band with $C=1$, independent of layer number, in contrast to our twisted 1+$n$ systems where the Chern number scales with the rhombohedral block thickness. We note that in our case, the mono/multi-RhG twist interface provides a well-defined moiré potential free from the sublattice ambiguity of hBN, allowing a straightforward comparison with theoretical calculations. In fact, the observed layer-dependent Chern hierarchy is in agreement with predictions of single-particle band model calculations, which remain qualitatively unchanged after introduction of Coulomb interactions at a mean-field level. These findings highlight the distinct roles of moiré potential and interlayer hybridization in determining topological band structures in graphene-based systems.

Moving forward, it is promising to explore smaller twist angles in $(1+n)$ systems, where the flatter band can promote correlation effect favorable for fractional states to emerge (Extended Data~\ref{fig:ED9}). Future extensions to multilayer configurations such as 2+3, 2+4, or 2+5 rhombohedral graphene will be particularly intriguing, as bilayer or thicker rhombohedral graphene can host isolated flat bands tunable by displacement field, providing a promising route to further enhance correlations in the topological bands~\cite{phong2025coulomb}. These systems could offer an ideal setting for realizing fractional states in high-Chern-number bands. Furthermore, given the plethora of unconventional superconductivity observed in rhombohedral graphene~\cite{seo2025family,kumar2025superconductivity,morissette2025superconductivity,nguyen2025hierarchy,deng2025superconductivity,kumar2025pervasive,guo2025flat}, creating lateral junctions to couple high-Chern-number insulators with superconductivity could open new opportunities for non-Abelian statistics and topological superconductivity~\cite{choi2025superconductivity,rontynen2015topological}.

\textit{Note: During the preparation of this manuscript, we became aware of a related study on twisted 1+5 rhombohedral graphene~\cite{liu2025diverse}.}

\bibliography{bibliography}

\newpage

\newpage

\newpage

\newpage

\ifshowfullcontent


\clearpage
\section*{Extended Data Tables}

\begingroup
\renewcommand{\thetable}{Table \arabic{table}}
\renewcommand{\tablename}{Extended Data}
\setcounter{table}{0}

\begin{table}[ht!]

\fontsize{10.5pt}{10.5pt}\selectfont
\caption{\textbf{Device summary.} AHE: anomalous Hall effect. $^\#$: Assuming an hBN dielectric constant $\epsilon_\mathrm{hBN}=3$ for the device. For the other devices, $\epsilon_\mathrm{hBN}$ is extracted from Landau fan features. Device (1+5)-1 was obtained by etching the device previously measured as (1+5)*-1 and (1+5)*-2; ``-1'' and ``-2'' denote different contact pairs in device (1+5)*. $B_\mathrm{c}$ denotes the critical field at which $R_\mathrm{yx}$ crosses zero. $^\dagger$: All other values were measured at 300 mK, except those marked here, which were measured at 50 mK.}
\vspace{0.5em}
\setlength{\tabcolsep}{4pt}
\begin{tabular}{|l|c|c|l|c|c|c|l|}

\hline
\hspace{2pt}Device & AHE & Gate & Twist ($^\circ$) & \makecell[c]{Best $R_\mathrm{yx}$ value (k$\Omega$) and \\ quantization percentage} & $B_\mathrm{c}$ (mT) & \makecell[c]{$\nu$, $D$ (V/nm) range \\ (>90\% quantization)} \\
\hline

(1+3)-1 & \ding{51}  & Gr& \hspace{8pt}1.28 
& \makecell{2.5 (0 T, 29\%) \\ 8.7 (4 T, quantized)} 
& 15 
& \makecell[l]{$\nu$: 1.23 $\sim$ 1.39, $D: -0.53$ \\ \hspace{53pt}(4 T)} \\
\hline

(1+3)-2 & \ding{51} & Gr& \hspace{8pt}1.33 
& 6.7 (0 T, 78\%) 
& 36 
& NA \\
\hline

(1+3)-3 & \ding{51} & Au & \hspace{8pt}1.11$^{\#}$ 
& 8.1 (0 T, 94\%) 
& 1 
& \makecell[l]{$\nu$: 2.57 $\sim$ 2.79, $D: -0.20$ \\ $\nu$: 2.61, $D: -0.22$ $\sim$ $-0.19$}\\
\hline

(1+3)-4 & \ding{55} & Gr& \hspace{8pt}0.94 
& NA & NA & NA \\
\hline

(1+3)-5 & \ding{55} & Gr& \hspace{8pt}0.90 
& NA & NA & NA \\
\hline

(1+4)-1 & \ding{51} & Gr & \hspace{8pt}1.34 
& 6.1 (0 T, 95\%) 
& 17 
& \makecell[l]{$\nu$: 1.04 $\sim$ 1.11, $D: -0.65$ \\ $\nu$: 1.05, $D: -0.65$ $\sim$ $-0.56$\\(gate range limited)}\\
\hline

(1+4)-2 & \ding{51} & Au& \hspace{8pt}1.39 
& \makecell{2.8 (0 T, 43\%) \\ 5.7 (3 T, 87\%)} 
& 5 
& NA \\
\hline

(1+4)-3 & \ding{51} & Gr & \hspace{8pt}1.22
& 4.8 (0 T, 74\%)
& 19 
& NA \\
\hline

(1+5)-1 & \ding{51} & Gr& \hspace{8pt}1.44 
& 5.1 (0 T, 99\%) 
& 3 
& \makecell[l]{$\nu$: 1.01 $\sim$ 1.09, $D: -0.61^\dagger$ \\ $\nu$: 1.02, $D: -0.65$ $\sim$ $-0.57^\dagger$} \\
\hline

(1+5)$^{*}$-1 & \ding{51} & Gr& \hspace{8pt}1.31 
& 4.8 (0 T, 93\%) 
& 22 
& \makecell[l]{$\nu$: 1.01 $\sim$ 1.04, $D: -0.58$ \\ $\nu$: 1.01, $D: -0.62$ $\sim$ $-0.57$}\\
\hline

(1+5)$^{*}$-2 & \ding{51} & Gr& \hspace{8pt}1.39 
& 4.8 (0 T, 93\%) 
& \makecell[c]{not \\measured} 
& \makecell[l]{$\nu$: 0.74 $\sim$ 1.02, $D: -0.73$ \\ $\nu$: 1.02, $D: -0.74$ $\sim$ $-0.71$}\\

\hline
\end{tabular}%
\label{tab:ED1}
\end{table}

\endgroup

\clearpage
\section*{Extended Data Figures}

\begingroup
\renewcommand{\thefigure}{Fig. \arabic{figure}}
\renewcommand{\figurename}{Extended Data}
\setcounter{figure}{0}

\begin{figure}[ht!]
    \centering
    \includegraphics[width=1\linewidth]{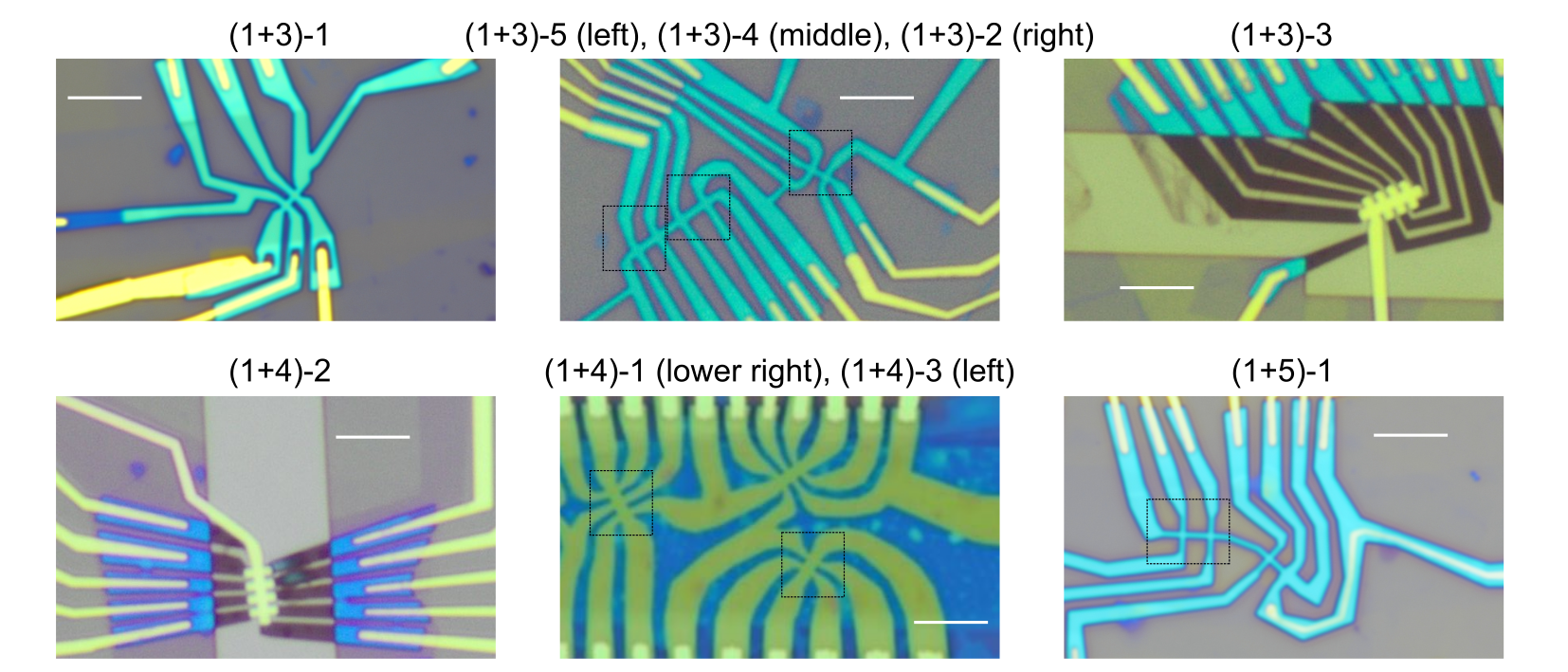}
    \caption{\textbf{Optical micrograph of fabricated $(1+n)$ graphene devices.} Scale bar, 5 µm.}
    \label{fig:ED1}
\end{figure}

\begin{figure}[ht!]
    \centering
    \includegraphics[width=0.9\linewidth]{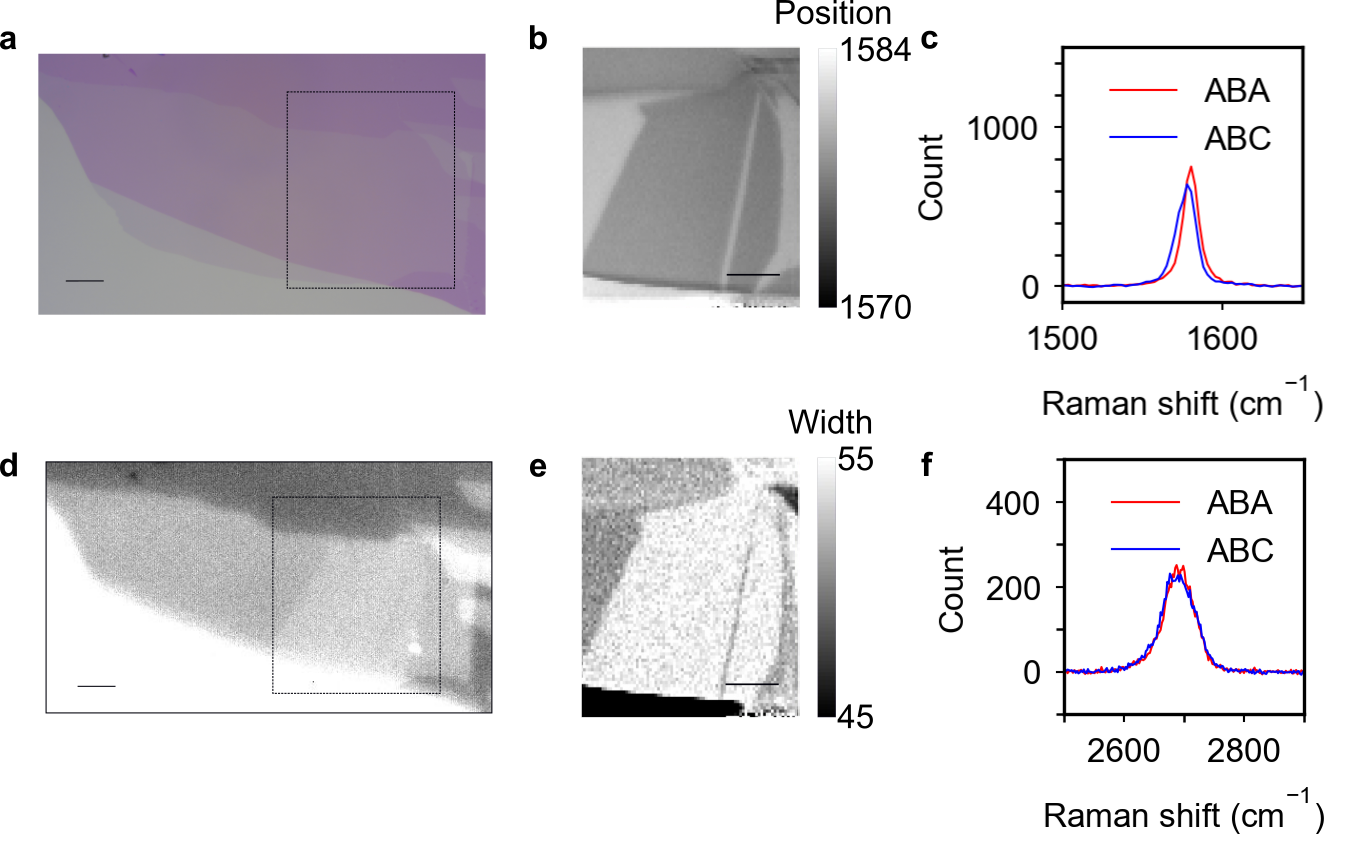}
    \caption{\textbf{Identification of rhombohedral stacking in multilayer graphene.} \textbf{a,} Optical micrograph of an exfoliated trilayer graphene flake. \textbf{b,} Spatial Raman map of the G-mode peak position over the region outlined in \textbf{a}, distinguishing ABA (Bernal, bright) and ABC (rhombohedral, dark) stacking domains. \textbf{c,} Representative Raman spectra of the G-mode peak for ABA (red) and ABC (blue) regions, showing a clear shift in peak position. \textbf{d,} Infrared image of the same flake, distinguishing ABA (dark) and ABC (bright) stacking domains. \textbf{e,} Spatial Raman map of the 2D peak width over the region outlined in \textbf{a}. \textbf{f}, Representative Raman spectra of the 2D band for ABA (red) and ABC (blue) domains, confirming rhombohedral stacking by the broader 2D peak width. Scale bar, 5 µm.}
    \label{fig:ED2}
\end{figure}

\begin{figure}[ht!]
    \centering
    \includegraphics[width=1\linewidth]{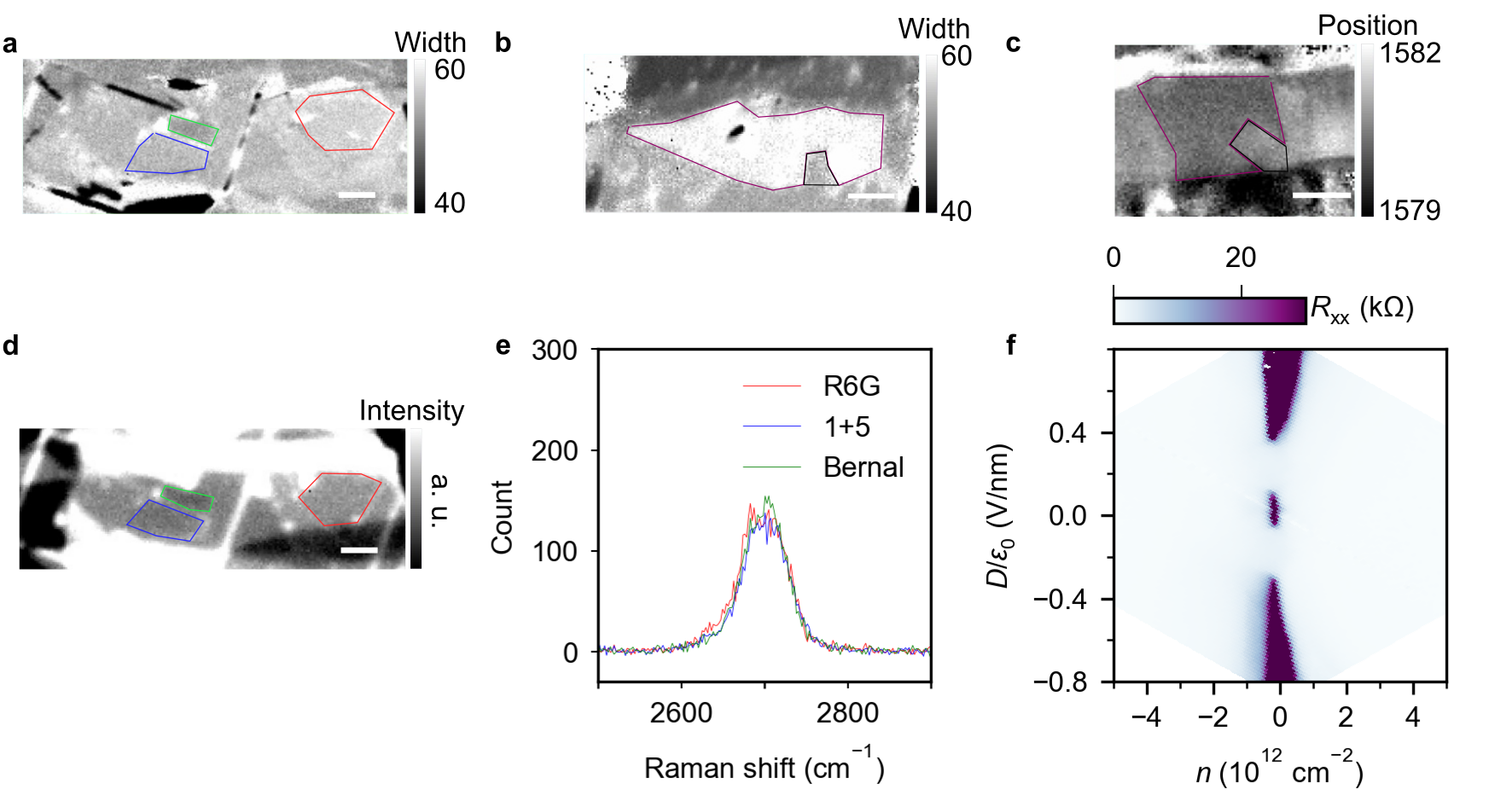}
    \caption{\textbf{Identification of stacking order in $(1+n)$ graphene stack.} Spatial Raman map of the 2D peak width (\textbf{a}) and Infrared image (\textbf{d}) of the $(1+5)$ graphene stack. Regions with different stacking orders are outlined with different colors. Red (brightest): rhombohedral hexalayer graphene. Blue (intermediate brightness): twisted (1+5) rhombohedral graphene. Green (darkest): Bernal graphene. Spatial Raman map of the 2D peak width in $(1+4)$ (\textbf{b}) and spatial Raman map of the G-mode peak position in $(1+3)$ (\textbf{c}) graphene stack. Purple: twisted (1+4) or (1+3) rhombohedral graphene. Black: Bernal graphene. \textbf{e}, Representative Raman spectra of different stacking domains in \textbf{a}. \textbf{f,} $R_\mathrm{xx}$ as a function of $\nu$ and $D/\varepsilon_{0}$ measured at the rhombohedral hexalayer graphene region in \textbf{a} and \textbf{d}. Scale bar, 5 µm.}
    \label{fig:ED3}
\end{figure}

\begin{figure}[ht!]
    \centering
    \includegraphics[width=0.95\linewidth]{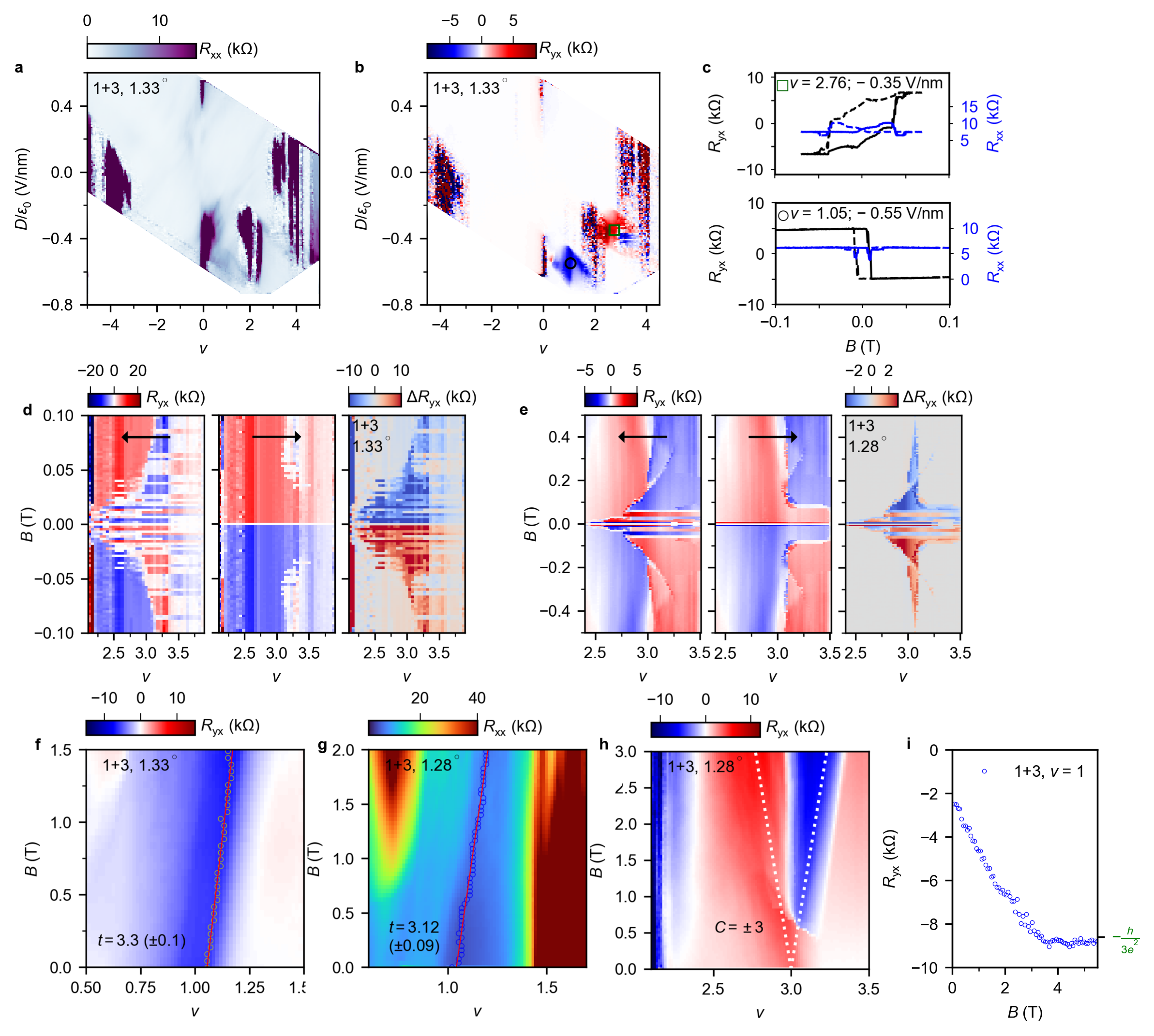}
    \caption{\textbf{Additional data of $(1+3)$ graphene devices at around $1.3^{\circ}$.} 
\textbf{a,} $R_\mathrm{xx}$ as a function of $\nu$ and $D/\varepsilon_{0}$ at $\theta=1.33^{\circ}$. \textit{B} = 0.06 T. \textbf{b,} Corresponding $R_\mathrm{yx}$ map at \textit{B} = 0.06 T. \textbf{c,} $R_\mathrm{yx}$ (black) and $R_\mathrm{xx}$ (blue) measured as a function of $B$ for selected values of $\nu$ and $D/\varepsilon_0$, which are labeled with corresponding markers in \textbf{b}. Solid (dashed) lines denote forward (backward) \textit{B} sweeps. \textbf{d,} Doping-induced magnetization switching. $\nu$ scanned backward (left), forward (middle), and the difference between forward and backward scans (right) at $D/\varepsilon_{0} = -0.36~\mathrm{V/nm}$. \textbf{e,} Doping-induced magnetization switching in the $1.28^{\circ}$-$(1+3)$ device at $D/\varepsilon_0$ = $-0.37$ V/nm. \textbf{f,} $R_\mathrm{yx}$ as a function of $\nu$ and \textit{B} for $1.33^\circ$-$(1+3)$ device. \textbf{g,} $R_\mathrm{xx}$ as a function of $\nu$ and \textit{B} for $1.28^\circ$-$(1+3)$ device. Red solid line is the linear fitting of the hollow circles which mark $\nu$ for the local minimum at each \textit{B}. \textbf{h,} $R_\mathrm{yx}$ as a function of $\nu$ and \textit{B} for $(1+3)$ device at $D/\varepsilon_0$ = $-0.29$ V/nm. The dotted lines which correspond to the noted Chern numbers are drawn as a guide to the eye. \textbf{i,} Linecut in main Fig. 4\textbf{a} along the dotted line from $\nu$ = 1. Data were all taken at \textit{T} = 0.3 K.}
    \label{fig:ED4}
\end{figure}

\begin{figure}[ht!]
    \centering
    \includegraphics[width=1\linewidth]{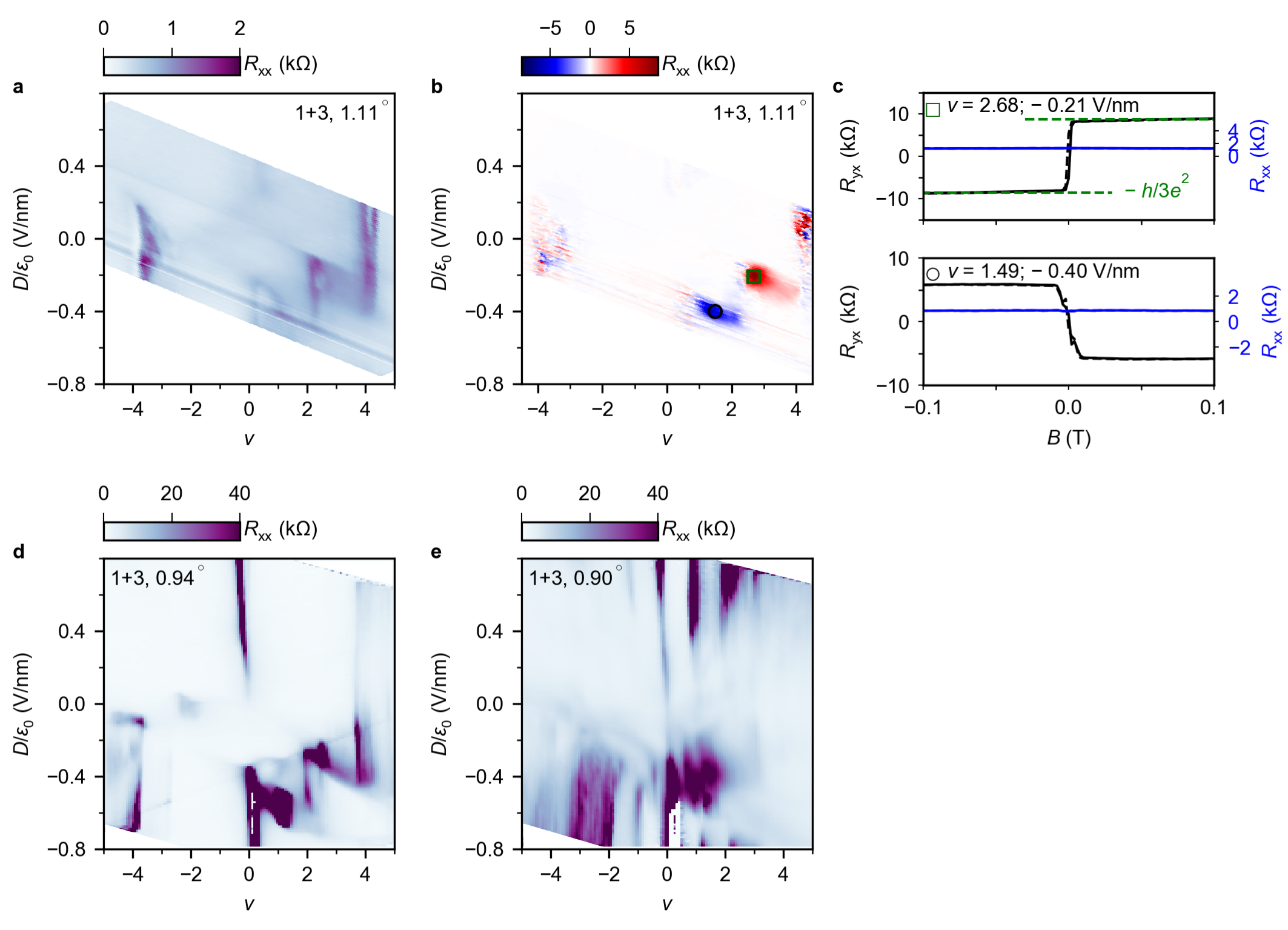}
    \caption{\textbf{Additional $(1+3)$ graphene devices at different twist angles.} \textbf{a,} $R_\mathrm{xx}$ as a function of $\nu$ and $D/\varepsilon_{0}$ at $\theta=1.11^{\circ}$. \textit{B} = 0.1 T. \textbf{b,} Corresponding $R_\mathrm{yx}$ map. \textit{B} = 0.1 T. \textbf{c,} $R_\mathrm{yx}$ (black) and $R_\mathrm{xx}$ (blue) measured as a function of $B$ for selected values of $\nu$ and $D/\varepsilon_0$, which are labeled with corresponding markers in \textbf{b}. Solid (dashed) lines denote forward (backward) \textit{B} sweeps. $|R_\mathrm{yx}|$ is quantized to $h/3e^2$ at $\nu=2.68$, in agreement with the Chern hierarchy. $R_\mathrm{xx}$ as a function of $\nu$ and $D/\varepsilon_{0}$ at $\theta=0.94^{\circ}$ (\textbf{d}) and $\theta=0.90^{\circ}$ (\textbf{e}) at \textit{B} = 0.06 T. Data were all taken at \textit{T} = 0.3 K.}
    \label{fig:ED5}
\end{figure}

\begin{figure}[ht!]
    \centering
    \includegraphics[width=1\linewidth]{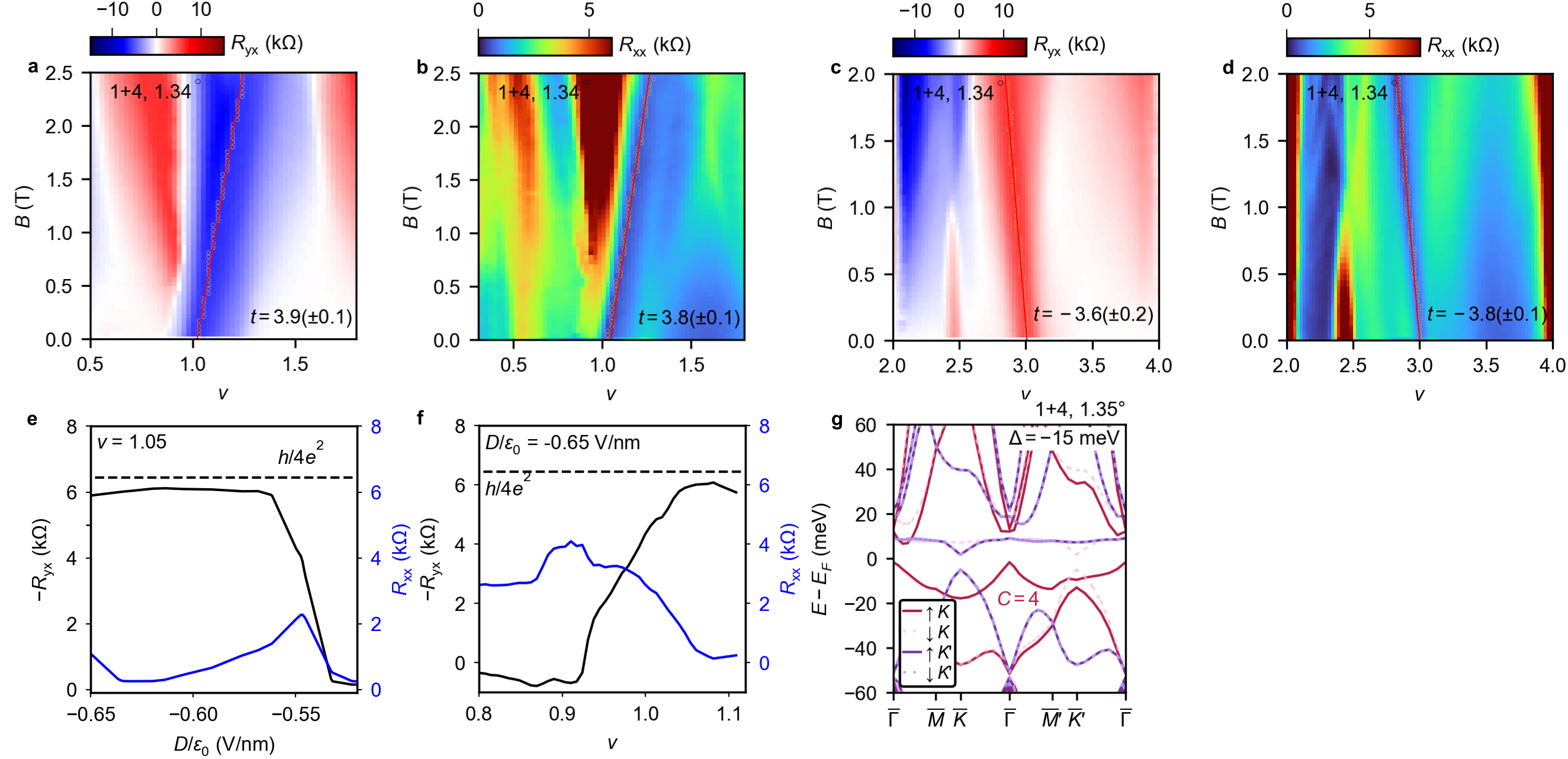}
    \caption{{\textbf{Additional data of $1.34^{\circ}-(1+4)$ graphene device.} $R_\mathrm{yx}$ (\textbf{a,c}) and $R_\mathrm{xx}$ (\textbf{b,d}) as a function of $\nu$ and \textit{B} for the $(1+4)$ device. \textit{T} = 0.3 K. Red solid lines are the linear fittings of the hollow circles which mark $\nu$ for the local extremum at each \textit{B}. \textbf{e,f,} Linecuts of $R_\mathrm{xx}$ and $-R_\mathrm{yx}$ at fixed $D/\epsilon_\mathrm{0}$ and $\nu$. \textit{B} = 0.2 T. \textit{T} = 0.3 K.} \textbf{g,} Hartree-Fock calculation of (1+4) system.}
    \label{fig:ED10}
\end{figure}

\begin{figure}[ht!]
    \centering
    \includegraphics[width=0.95\linewidth]{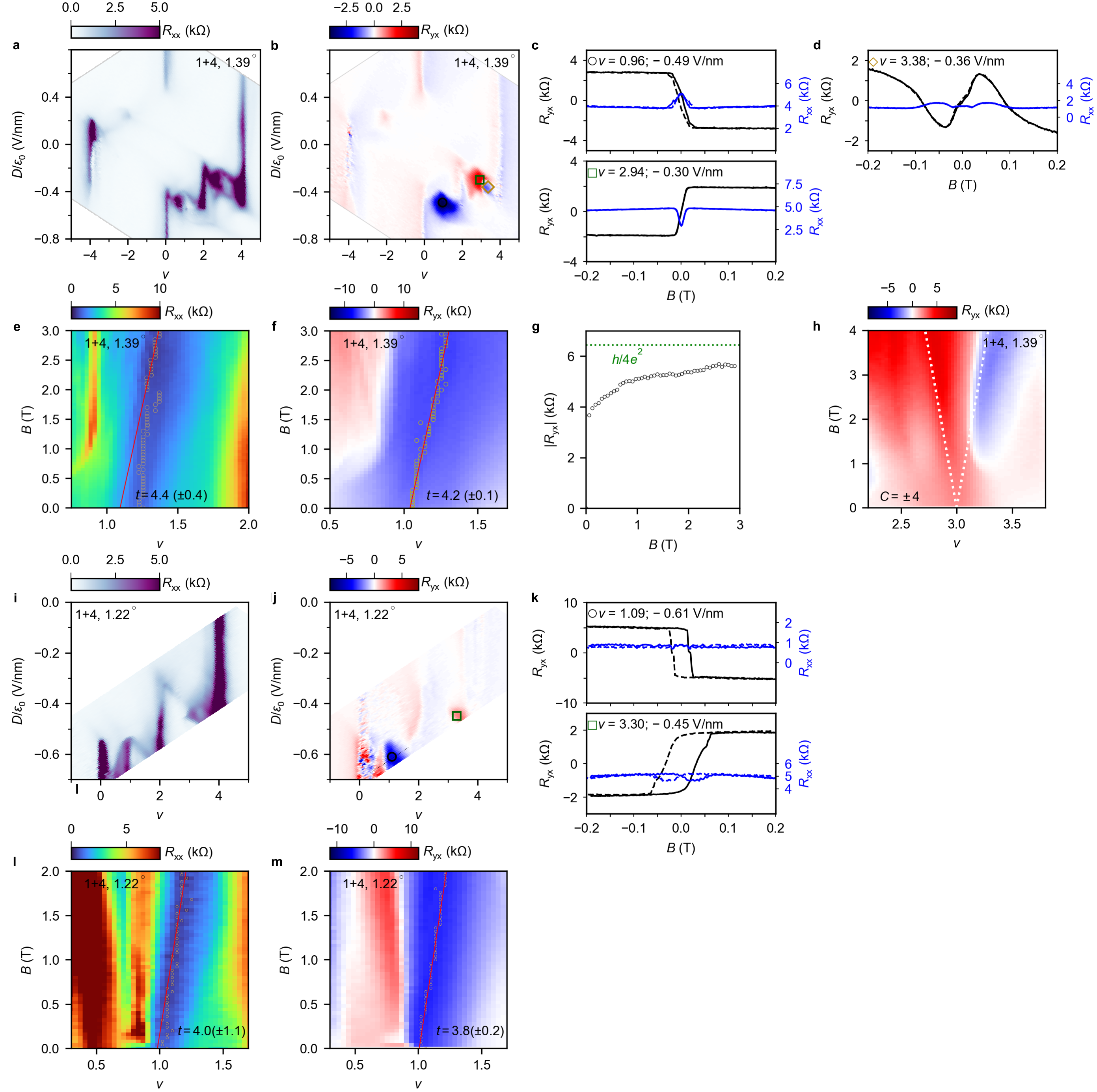}
    \caption{\textbf{Additional data of $(1+4)$ graphene devices.} Dual-gate maps of $R_\mathrm{xx}$ and $R_\mathrm{yx}$ of $1.39^{\circ}-(1+4)$ device (\textbf{a,b}) and $1.22^{\circ}-(1+4)$ device (\textbf{i,j}). \textit{B} = 0.2 T.
    $R_\mathrm{yx}$ and $R_\mathrm{xx}$ measured as a function of $B$ (\textbf{c,d} for $1.39^{\circ}$ device, \textbf{k} for $1.22^{\circ}$ device). Solid (dashed) lines denote forward (backward) \textit{B} sweeps. 
    $R_\mathrm{xx}$ or $R_\mathrm{yx}$ as a function of $\nu$ and \textit{B} at $D/\varepsilon_0$ = $-0.50$ V/nm (\textbf{e,f}) and $-0.26$ V/nm (\textbf{h}). Red solid lines in \textbf{e,f} are the linear fittings of the hollow circles which mark $\nu$ for the local extremum at each \textit{B}. Fitting in \textbf{e} is performed for data above 2 T where a clear dispersion is developed.
    The dotted white line which correspond to the noted Chern number is drawn as a guide to the eye in \textbf{h}. 
    \textbf{g,} Linecut in \textbf{f} along $C=4$ trajectory from $\nu$ = 1. $|R_\mathrm{yx}|$ approaches $h/4e^2$ but without quantization, likely due to moiré inhomogeneity. 
    \textbf{l,m} $R_\mathrm{xx}$ and $R_\mathrm{yx}$ as a function of $\nu$ and \textit{B} taken along dashed line show in \textbf{j}.
    Data were all taken at \textit{T} = 0.3 K.}
    \label{fig:ED6}
\end{figure}

\begin{figure}[ht!]
    \centering
    \includegraphics[width=0.83\linewidth]{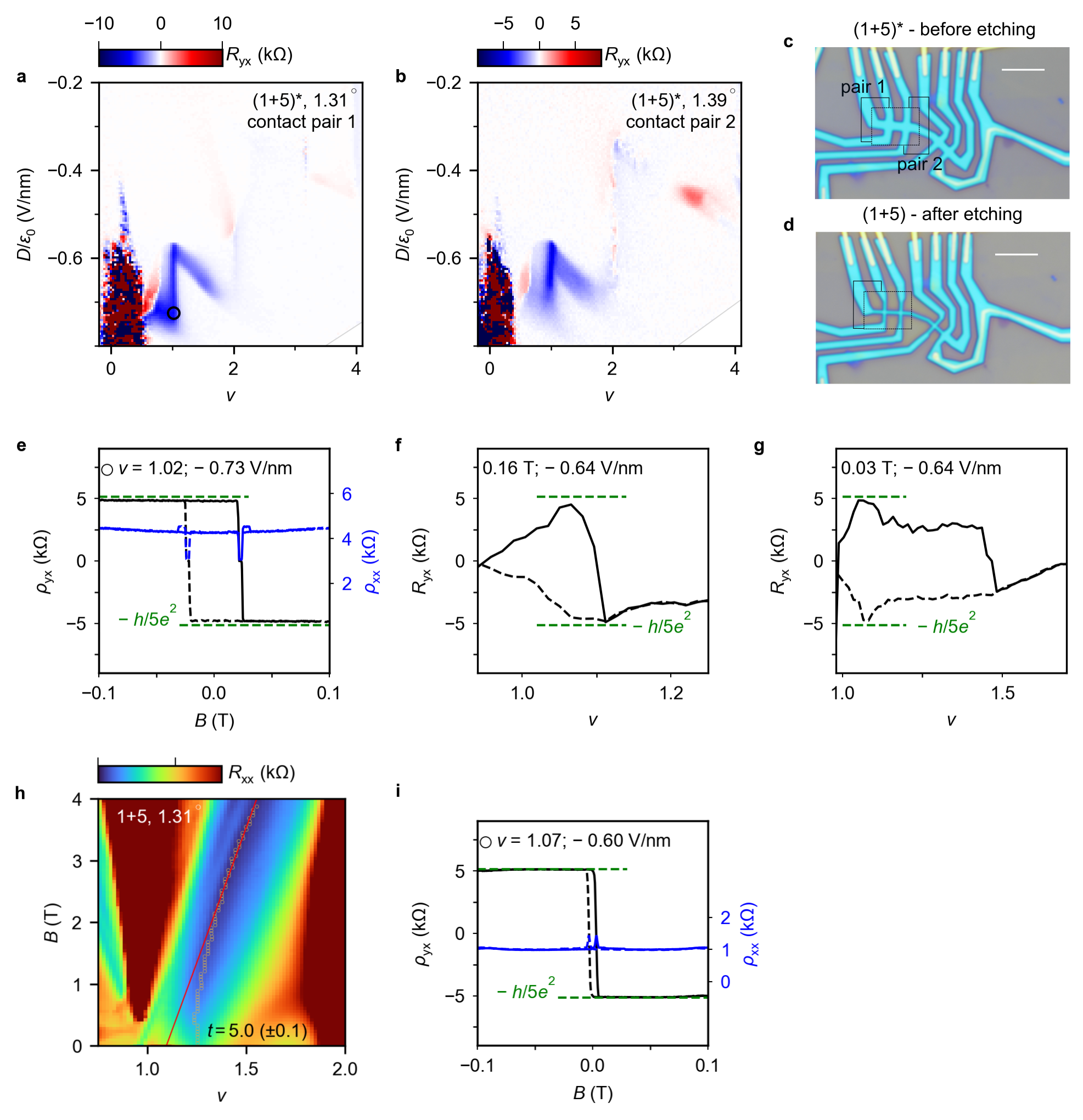}
    \caption{\textbf{$(1+5)$ graphene device before and after etching.} $R_\mathrm{yx}$ as a function of $\nu$ and $D/\varepsilon_{0}$ for two contact pairs before etching (\textbf{a,b}). \textit{B} = 0.06 T, \textit{T} = 0.3 K. Optical micrograph of device before (\textbf{c}, device (1+5)*) and after (\textbf{d}, device (1+5)) etching. The contact pair used for measuring $R_\mathrm{yx}$ in device (1+5) is denoted in \textbf{d}. Scale bar, 5 µm. \textbf{e,} $\rho_\mathrm{yx}$ (black) and $\rho_\mathrm{xx}$ (blue) measured as a function of $B$ at around $\nu=1$, labeled with the corresponding marker in \textbf{a}. Solid (dashed) lines denote forward (backward) \textit{B} sweep. \textit{T} = 0.3 K. Magnetotransport measurement of the same device at a different displacement field is shown in Fig. 4b,c in the main text. \textbf{f,g,} Doping-induced switching of high-Chern-number magnetization in device (1+5). Solid (dashed) lines denote forward (backward) sweeps. Data over a more complete range of magnetic fields are shown in Fig. 4f of the main text. \textit{T} = 0.05 K. \textbf{h,} $R_\mathrm{xx}$ as a function of $\nu$ and \textit{B} for $1.31^\circ$-$(1+5)$ device. $D = -0.66$ V/nm. \textbf{i,} $\rho_\mathrm{yx}$ (black) and $\rho_\mathrm{xx}$ (blue) measured as a function of $B$ at around $\nu=1$ in $1.44^\circ$-(1+5) device. $R_\mathrm{yx}$ and $R_\mathrm{xx}$ data are shown in Fig. 3f. }
    \label{fig:ED7}
\end{figure}

\begin{figure}[ht!]
    \centering
    \includegraphics[width=0.8\linewidth]{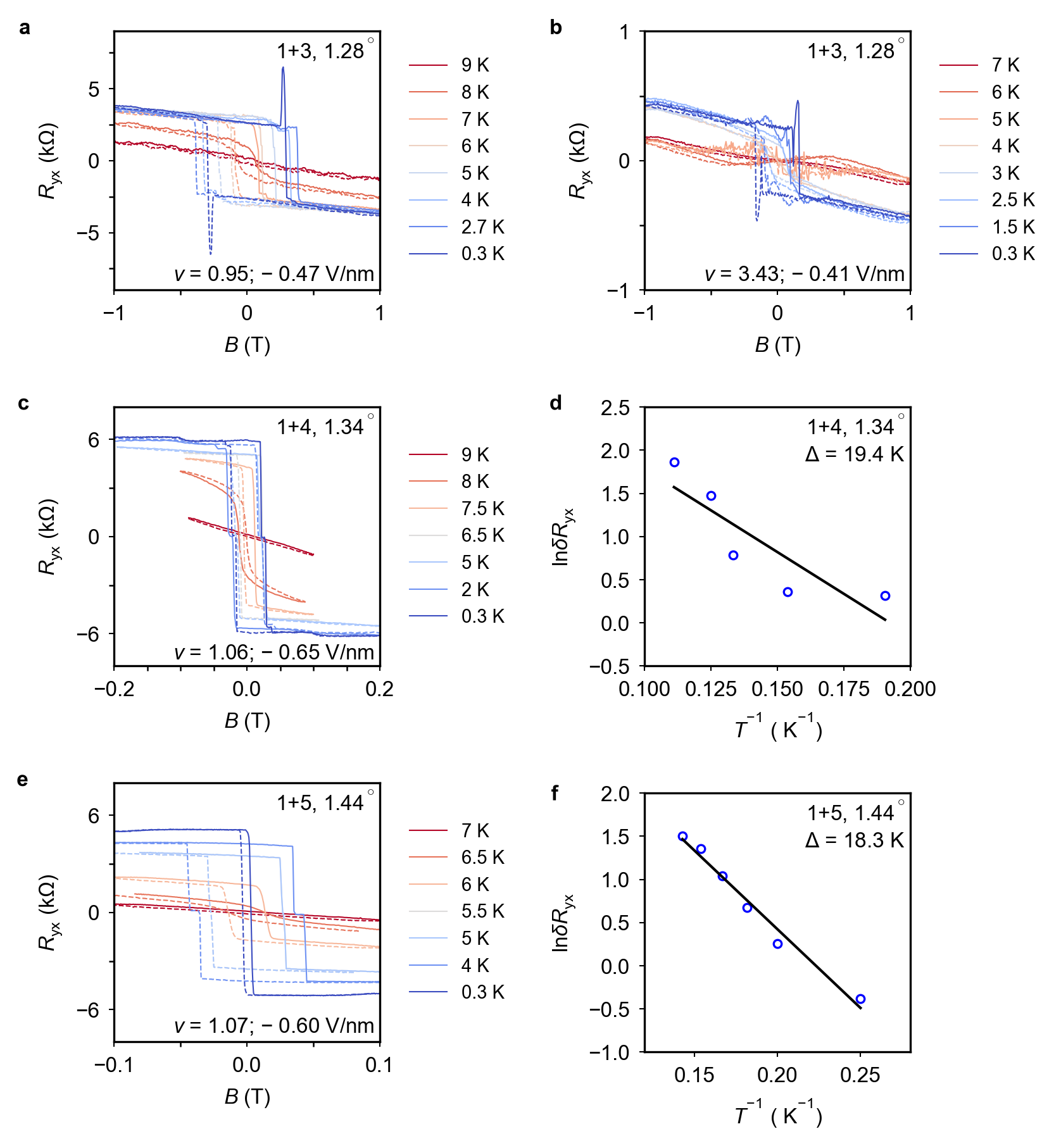}
    \caption{\textbf{Temperature dependence of anomalous Hall effect in $(1+n)$ systems.} $R_\mathrm{yx}$ measured as a function of $B$ for selected values of $\nu$ and $D/\varepsilon_0$. Solid (dashed) lines denote forward (backward) \textit{B} sweeps. \textbf{a,b,} $(1+3)$ device. Room-temperature RC filters were used, resulting in a larger coercive field than in Fig. 3d in the main text. \textbf{c,} $(1+4)$ device. \textbf{e,} $(1+5)$ device. A few measurements were taken between the 0.3 K data and other data, which probably led to a change in the domain pinning configuration and resulted in a smaller hysteresis loop at 0.3 K. \textbf{d,f,} Deviation of $R_\mathrm{yx}$ from quantization plotted as a function of temperature, extracting a gap size $\Delta=19.4$ K for (1+4) device and $\Delta=18.3$ K for (1+5) device by fitting $\delta R_\mathrm{yx}=e^{-\Delta/k_BT}$.}
    \label{fig:ED8}
\end{figure}

\begin{figure}[ht!]
    \centering
    \includegraphics[width=1\linewidth]{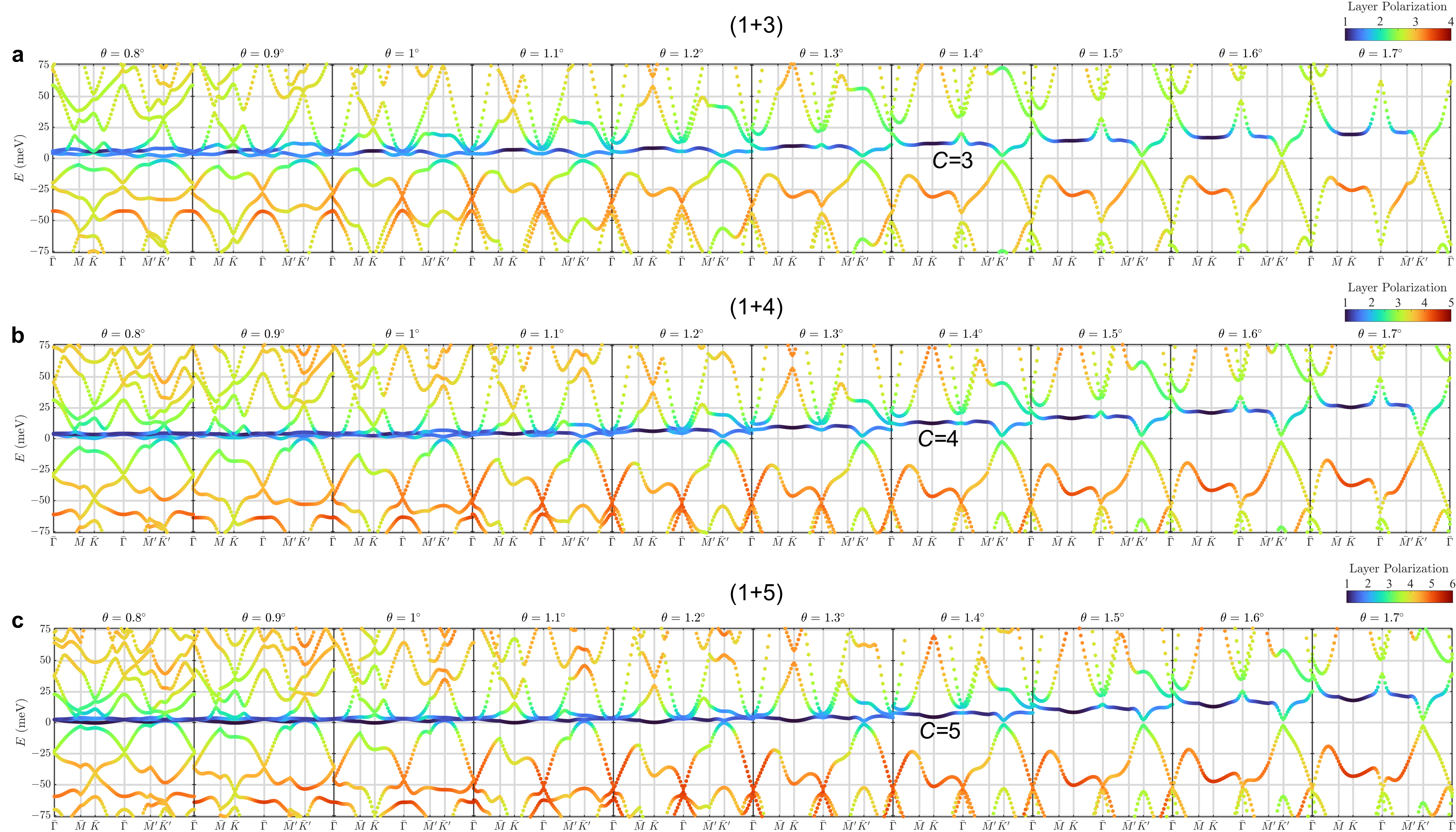}
    \caption{\textbf{Evolution of non-interacting band structures for $(1+n)$ systems as a function of angles.} \textbf{a,} Band structure of $(1+3)$ configuration at $\Delta = -20$ meV. \textbf{b,} Band structure of $(1+4)$ configuration at $\Delta = -20$ meV. \textbf{c,} Band structure of $(1+5)$ configuration at $\Delta = -15$ meV. Each energy eigenstate is color coded by its layer polarization. }
    \label{fig:ED9}
\end{figure}

\endgroup


\clearpage
\section{Methods}
\small{
\subsection{Flake identification}
hBN and graphite crystals were exfoliated onto SiO$_2$ (285 nm-295 nm)/Si substrates. The thickness of hBN flakes was acquired by atomic force microscopy (AFM), and the thickness of the 1-5 layers of graphene flakes were identified by optical contrast. The graphene stacking order was identified using an infrared (IR) camera as described in Ref~\cite{lu2025extended}, and then mapped with Raman spectroscopy (Renishaw Invia Reflex Micro Raman, wavelength 532 nm) to confirm the stacking order (Extended Data~\ref{fig:ED2}). Then it is cut by AFM anodic cutting to isolate the rhombohedral stacking pieces as described in Ref~\cite{li2018electrode}. The crystallographic axis of rhombohedral graphene and some of the hBNs were identified using lateral force microscopy (LFM). 

\subsection{Transfer processes}
We first prepared the bottom stacks. For devices (1+3)-3 and (1+4)-1, we evaporated PdAu(13 nm)/Ti(2 nm) on SiO$_2$(285 nm)/Si substrates by thermal evaporation for the bottom electrodes. After heat annealing at 300 °C under a mixture of H$_2$ and Ar gases, we transferred hBN as a bottom gate dielectric using poly(bisphenol A carbonate) (PC) film on Polydimethylsiloxane (PDMS). For the other devices, an hBN and a graphite flake were sequentially picked up and then released to a SiO$_2$/Si substrate at 170 °C. After the PC was dissolved, contact mode AFM was performed at a deflection voltage of 0.1$\sim$0.2 V for cleaning the PC residues. Top stacks for device (1+3)-3 and (1+4)-1 were made by the sequential pickup of top hBN and twisted mono-multilayer graphene. Top stacks for the other devices were made by the sequential pickup of topmost hBN, top graphite, top hBN, and twisted mono-multilayer graphene. Twisted mono-multilayer graphene was picked up with the cut-and-stack method described in Ref.~\cite{park2021flavour}. The whole stack was then released to the bottom stack at between 110 and 170 °C.  During the transfer, we intentionally misalign the graphene from top and bottom hBNs based on the LFM of graphene and the straight edges or the LFM of the hBNs.

\subsection{Post-transfer fabrication}
Raman spectroscopy and IR camera were used again to search for rhombohedral regions survived after the transfer. After the transfer, we sometimes identify three types of regions, corresponding to the rhombohedral flake survived with the monolayer twisted on top, rhombohedral flake relaxed to Bernal stacking, and the rhombohedral flake survived but with the monolayer fully aligned and forming the $(1+n)$ layers of rhombohedral graphene. They typically show different contrast in the Raman spectroscopy and the IR images (Extended Data~\ref{fig:ED3}), but require a careful comparison between different regions to identify the exact stacking orders, and sometimes require transport measurement to finally confirm the stacking order. After Raman scanning and IR imaging, we identify the stack with an optical microscope and AFM for bubble-free regions. Then the stack was etched into a Hall bar shape by reactive ion etching. All the contacts (and metal top gates) were deposited with Au/Cr with a thermal evaporator.

\subsection{Lateral force microscopy measurements }
To determine the crystallographic axis, we performed LFM measurements on the rhombohedral graphene flakes and some of the top and bottom hBN flakes. To avoid change of the stacking order by the LFM, we scanned on the Bernal stacking part that is from the same piece of graphite as the rhombohedral graphene. The LFM measurements were performed with Asylum Research Cypher S atomic force microscope at room temperature. We used the Asyelec-01-R2 tip with a force constant of around 2.8 N m$^{-1}$ and a deflection setpoint of around 2 V.

\subsection{Transport measurements}
The devices were bonded by aluminum wire. The four-probe measurements were done using lock-in amplifiers (SRS: SR830 and SR860), a current preamplifier (DL: Model 1211) and voltage preamplifiers (SRS: SR560 and Ithaco: Model 1201) at the frequency of 17$\sim$35 Hz. The gate voltages were applied by source meters (Keithley: Model 2400 and 2450). Devices were measured in a He-3 cryostat (Janis Research) or a dilution fridge (Leiden Cryogenics).

\subsection{Density calibration}
We calculate the carrier density using a parallel-plate capacitor model, $n = \epsilon_0 (\epsilon_\mathrm{t} V_\mathrm{t}/d_\mathrm{t} + \epsilon_\mathrm{b} V_\mathrm{b}/d_\mathrm{b})/e$, where $V$, $\epsilon$, and $d$ denote the gate voltage, dielectric constant, and hBN thickness, respectively. The subscripts “t” and “b” denote the top and bottom gate parameters. The hBN thicknesses are measured using atomic force microscopy.

The dielectric constant can then be determined by tracking Landau level features in dual-gate maps at finite $B$. The change in carrier density between adjacent Landau levels corresponds to $\Delta n = g e B/h$, where the degeneracy $g = 4$ in the full metal phase. By measuring the spacing in $V_\mathrm{t}$ and $V_\mathrm{b}$ between adjacent Landau levels, we can fit the density equation above and extract $\epsilon_\mathrm{t}$ and $\epsilon_\mathrm{b}$.


Importantly, we note that the Chern number extracted from the Středa relation, $\Delta n = C e \Delta B/h$, can be determined by direct comparison with the Landau level spacings in the full metal phase as described above, without the need to calculate the dielectric constants. Therefore, the inferred hierarchy $C = n$ is robust against variations in density calibration. The uncertainty in the density calibration only affects our estimation of the twist angle, which is calculated using the carrier density at full filling $n_0 = (8/\sqrt{3}a^2)\,\theta^2$, where \textit{a} is the lattice constant.

\subsection{Magnetic hysteresis analysis}
In the manuscript, we apply standard symmetrization and antisymmetrization procedures to the magnetic hysteresis data, to eliminate mixing between $R_{xx}$ and $R_{yx}$ caused by contact misalignment (b (f) for backward (forward) \textit{B} sweeps):
\begin{align*}
R_{\mathrm{xx}}^{\mathrm{sym},\,\mathrm{f}}(B)
&=
\frac{
R_{\mathrm{xx}}^{\mathrm{raw},\,\mathrm{f}}(B)
+
R_{\mathrm{xx}}^{\mathrm{raw},\,\mathrm{b}}(-B)
}{2},
R_{\mathrm{xx}}^{\mathrm{sym},\,\mathrm{b}}(B)
=
R_{\mathrm{xx}}^{\mathrm{sym},\,\mathrm{f}}(-B);
\\[2pt]
R_{\mathrm{yx}}^{\mathrm{antisym},\,\mathrm{f}}(B)
&=
\frac{
R_{\mathrm{yx}}^{\mathrm{raw},\,\mathrm{f}}(B)
-
R_{\mathrm{yx}}^{\mathrm{raw},\,\mathrm{b}}(-B)
}{2},
R_{\mathrm{yx}}^{\mathrm{antisym},\,\mathrm{b}}(B)
=
-\,R_{\mathrm{yx}}^{\mathrm{antisym},\,\mathrm{f}}(-B).
\end{align*}

\subsection{Středa analysis procedure}

The Chern number is obtained from the Středa relation $C = (h/e)\partial n/\partial B$, where the carrier density $n$ is determined from the Landau fan analysis. In practice, because $R_\mathrm{yx}$ directly reflects the topologically protected Hall response while $R_\mathrm{xx}$ can be more sensitive to disorder and dissipation~\cite{weis2011metrology}, we primarily track the dispersing feature in $R_\mathrm{yx}$ and use $R_\mathrm{xx}$ as a consistency check.

For the devices where clear dispersing features are observed, we take line cuts on the $R_\mathrm{yx}$ map along the integer-$C$ trajectory that best follows the prominent dispersing feature and confirm the \textit{C} number by observing the quantization, as has been done for (1+3) and (1+5) devices in the main text.

For the devices that do not reach full quantization, we track the dispersing maxima in $R_\mathrm{yx}$ within a window of $\Delta\nu=0.5$ and perform linear fitting to extract the Chern number, which is done for the (1+4) device and the additional (1+3) device in Extended Data Fig. 4 and 6. The uncertainty of the slope is estimated from the linear regression fitting error.

As a robustness check to avoid uncertainty or outliers in the identification of local extrema location, we also determine the local extrema location using a weighted average of the five nearest data points. For $R_\mathrm{xx}$ minima ($R_\mathrm{yx}$ maxima), the weights are taken proportional to $-R_\mathrm{xx}$ ($|R_\mathrm{yx}|$), so that points closer to the extremum contribute more strongly. This procedure yields slopes that differ by less than $\sim0.1$ in the extracted Chern number compared with using a single extremum point.

\section{Acknowledgments}
We thank P. Ledwith and J.M. Park for fruitful discussions, and L-Q.X. for the help in the measurement setup. This research was supported by the Center for the Advancement of Topological Semimetals, an Energy Frontier Research Center funded by the U.S. Department of Energy Office of Science, through the Ames Laboratory under contract DE-AC02-07CH11358 (measurements and data analysis), the Air Force Office of Scientific Research (AFOSR) grant FA9550-21-1-0319 (characterization), the Office of Naval Research (ONR) grant N000142412440 (device fabrication), the MIT/Microsystems Technology Laboratories Samsung Semiconductor Research Fund, the Gordon and Betty Moore Foundation’s EPiQS Initiative through grant GBMF9463, and the Ramon Areces Foundation. This work was performed in part at the Harvard University Center for Nanoscale Systems (CNS), a member of the National Nanotechnology Coordinated Infrastructure Network (NNCI), which is supported by the National Science Foundation under NSF ECCS award No. 1541959. This work was carried out in part through the use of MIT.nano’s facilities. X.W. acknowledges the support from Mathworks fellowship. L.A.B. gratefully acknowledges the generous support of the Mauricio and Carlota Botton Foundation. K.W. and T.T. acknowledge support from the JSPS KAKENHI (Grant Numbers 21H05233 and 23H02052), the CREST (JPMJCR24A5), JST and World Premier International Research Center Initiative (WPI), MEXT, Japan. Numerical calculations are done using the High Performance Compute Cluster  of the Research Computing Center (RCC) at Florida State University. V.T.P. and C.L. are supported by  start-up funds from Florida State University and the National High Magnetic Field Laboratory. The National High Magnetic Field Laboratory is supported by the National Science Foundation through NSF/DMR-2128556 and the State of Florida. 

\section{Author contributions}
X.W. and L.A.B. conceived the project. X.W., L.A.B., and S.R. fabricated the devices and performed measurements with the help of W.I.C. V.T.P. performed the theoretical calculations supervised by C.L. K.W. and T.T. grew the
hBN crystals. P.J-H. supervised the project. X.W. and L.A.B. analyzed the data and wrote the manuscript with input from all the other authors.

\section{Competing interests}
The authors declare no competing financial interests.

\else

\renewcommand{\thetable}{Table \arabic{table}}
\renewcommand{\tablename}{Extended Data}

\renewcommand{\thefigure}{Fig. \arabic{figure}}
\renewcommand{\figurename}{Extended Data}

\setcounter{figure}{0}
\setcounter{table}{0}

\refstepcounter{table}
\phantomsection
\label{tab:ED1}

\refstepcounter{figure}
\phantomsection
\label{fig:ED1}

\refstepcounter{figure}
\phantomsection
\label{fig:ED2}

\refstepcounter{figure}
\phantomsection
\label{fig:ED3}

\refstepcounter{figure}
\phantomsection
\label{fig:ED4}

\refstepcounter{figure}
\phantomsection
\label{fig:ED5}

\refstepcounter{figure}
\phantomsection
\label{fig:ED6}

\refstepcounter{figure}
\phantomsection
\label{fig:ED7}

\refstepcounter{figure}
\phantomsection
\label{fig:ED8}

\refstepcounter{figure}
\phantomsection
\label{fig:ED9}

\fi

\end{document}